\documentclass[doublecol]{epl2}
\usepackage{amssymb}
\usepackage{graphicx}
\usepackage{booktabs}
\usepackage{tabularx}
\usepackage{array}
\usepackage{lscape}
\usepackage{longtable}
\usepackage{multirow}
\usepackage{amsmath}
\usepackage{colortbl}
\usepackage{algorithm}
\usepackage{algorithmic}

\title{A biologically inspired model for transshipment problem}
\shorttitle{title} 

\author{C. Gao\inst{1} \and C. Yan\inst{2,3}  \and D. Wei\inst{1,4} \and Y. Hu\inst{5} \and S. Mahadevan\inst{6} \and Y. Deng\inst{1}}
\shortauthor{F. Author \etal}

\institute{
  \inst{1} School of Computer and Information Science, Southwest University - Chongqing 400715, China\\
  \inst{2} University of Chinese Academy of Sciences - Beijing 100190, China\\
  \inst{3} Computer Network Information Center, Chinese Academy of Sciences - Beijing 100190, China\\
  \inst{4} School of Science, Hubei University for Nationalities, Enshi 445000, China\\
  \inst{5} Institute of Business Intelligence and Knowledge Discovery, Guangdong University of Foreign Studies, Guangzhou 510006, China
  \inst{6} School of Engineering, Vanderbilt University - TN 37235, USA
}

\abstract{Transshipment problem is one of the basic operational research problems. In this paper, our first work is to develop a biologically inspired mathematical model for a dynamical system, which is first used to solve minimum cost flow problem. It has lower computational complexity than \emph{Physarum Solver}. Second, we apply the proposed model to solve the traditional transshipment problem. Compared with the conditional methods, experiment results show the provided model is simple, effective as well as handling problem in a continuous manner.}

\begin{document}
\maketitle

\section{Introduction}
\emph{Physarum polycephalum} is a 'smart' true slime mold, which exhibits a various of pattern formation behaviors, such as growth, food foraging, nutrient transportation, propagation of pseudopodium and maintenance of tube shape. Recently, experiment results have shown \emph{Physarum} can construct biological transport networks efficiently~\cite{tero2010rules,gunji2008minimal,gunji2011adaptive,hu2013adaptation}, solve Shortest Path (SP) problem ~\cite{nakagaki2000intelligence,adamatzky2012slime,nakagaki2007minimum} and Steiner tree problem with three or four food sources~\cite{nakagaki2004smart}. By extracting from the complex behaviors of \emph{Physarum}, there have been some simple and distinct models developed for solving SP problem ~\cite{adamatzky2012slime,tero2007mathematical,ma2009current,nakagaki2007minimum} in a maze and other optimization problems so far~\cite{gunji2008minimal,hu2013adaptation}. These include a mathematical model \emph{Physarum Solver}~\cite{tero2007mathematical}, Oregonator model of the Belousov-Zhabotinsky (BZ) medium~\cite{adamatzky2009BZ,adamatzky2012slime}, particle-like agent model simulating the growth of \emph{Physarum}~\cite{jones2010emergence,jones2012emergence} and the cell model~\cite{gunji2008minimal,gunji2011adaptive}. These models have been applied to deal with some extended SP problems~\cite{zhangyajuan2013biologically,zhangxiaoge2013physarum,zhangxiaoge2013adaptive}, Spanning tree problem~\cite{adamatzky2009BZ}, road navigation~\cite{tero2006physarum,zhangxiaoge2013route}, efficient transportation networks~\cite{tero2010rules,adamatzky2012world,adamatzky2013bio-development} and other problems~\cite{zhangxiaoge2013knapsack,aono2011TSP,watanabe2011traffic,gaocai2013TP}.

As mentioned above, \emph{Physarum Solver}~\cite{tero2007mathematical} is capable of finding SP in a maze where two separate food sources (FSs) are placed at two exits, respectively. \emph{Physarum polycephalum} always prefers to transporting its nutrients and chemical signals in an effective manner. Nakagaki \emph{et al.} show that this simple organism has the ability to find the minimum-length path of a maze~\cite{nakagaki2000intelligence}. In other words, it can be interpreted as transporting a certain amount of flow among the minimum-length path. As a consequence, SP problem can be regarded as single-source single-sink minimum cost flow problem.

In transshipment problem, items are supplied from different sources to different destinations. It is sometimes economical if the shipment passes some transient nodes in between sources and destinations. The objective is to minimize the total cost of shipments. and thus the shipment passes through one or more intermediate nodes before it reaches its destinations~\cite{hoppe2000quickest,herer2006multilocation}. Therefore, transshipment problem is a special multi-source multi-sink minimum cost flow problem.

In this Letter, we first develop a biologically inspired mathematical model in response to local information (potential difference). This model has lower computational complexity than \emph{Physarum Solver}. Second, we extend the proposed model to solve transshipment problem in an continuous manner, which is different the discrete simplex method.

The paper is organized as follows. In Section 2 we present the biologically inspired mathematical model. Then, extend this model to solve transshipment problem. In Section 3 we discuss the numerical study, which illustrate the efficiency of the proposed model. Section 4 concludes the paper.
\section{Method}
\subsection{A biologically inspired mathematical model}
Let $G = (N,E,L)$ be a network, where $N$ is a set of nodes, $E$ is a set of directed edges and $L_{ij}$ is the length of edge $E_{ij}$. A particle enters the network from one of the source nodes $S_i(i=1,2,...,l_S)$ and moves out of one sink node $T_i(i=1,2,...,l_T)$. In this letter, we regard the number of particles as a metric of potential. Larger the number of particles, higher potential of node $i$. The particle always walks towards nodes with lower potential. It is similar to the natural phenomena that water move from one area to another due to water potential difference. The movement of numerous particles forms particle flow among tubular edge. Assume the flow along the tube edge is approximately Poiseuille flow. $Q_{ij}$ denotes the flux through $E_{ij}$. The flux $Q_{ij}$ is given as follows
\begin{equation}
\label{eqofpoiseuille}
{Q_{ij}} = {D_{ij}}\frac{{{\Phi _i} - {\Phi _i}}}{{{L_{ij}}}},
\end{equation}
where $D_{ij}$ is the conductivity of edge $E_{ij}$, $\Phi_i$ is the number of particles stored at node $i$. The direction of the flow is determined by the sign $\Phi_i - \Phi_j$. If $\Phi_i > \Phi_j$, the particle flow goes from node $i$ to node $j$., otherwise it goes from the opposite direction. According to Eq.~(\ref{eqofpoiseuille}), particles prefer to move along an edge with low $L/D$. Note that the length $L$ is a fixed constant. In order to make particles move along edges with low $L/D$ values, we update the conductivity $D_{ij}$ in response to particle flow as follows
\begin{equation}
\label{eqofevolution}
\frac{d}{{dt}}{D_{ij}} = {Q_{ij}} - {D_{ij}}
\end{equation}
We call Eq.~(\ref{eqofevolution}) \emph{evolution equation}. This equation implies that conductivity vanishes gradually if there is no flow along the edge. The initial conductivity of each edge is equal to a fixed positive number. The dynamics of the particles at node $i$ can be described as follows
\begin{equation}
\label{eqofparticle}
\frac{d}{{dt}}{\Phi _i} = \sum\limits_{e \in {E_i}} {{Q_e}}
\end{equation}
where $E_i$ is the set adjacent edges of node $i$. For this dynamic system, let the inflow of node $S_i$ be $I_{S(i)}$ and the outflow of sink node $T_i$ $I_{T(i)}$. Consider the conservation law, there is a constraint condition $\sum\limits_{i = 1}^{{l_S}} {{I_{S(i)}}}  = \sum\limits_{j = 1}^{{l_T}} {{I_{T(j)}}}$. Note that the sources of the network have injected particle flow so that particles at source nodes will eventually be at equilibrium.
\subsection{Algorithm implementation}
According to the model described above, its numerical scheme is given as follows.
We first solve \emph{evolution equation} (\ref{eqofevolution}) using a semi-implicit scheme
\begin{equation}
\label{eqDiscreteEvolution}
\frac{{{D_{ij}}(t + \Delta t) - {D_{ij}}(t)}}{{\Delta t}} = {Q_{ij}}(t) - {D_{ij}}(t + \Delta t)
\end{equation}
where $\Delta t$ is the time internal, $0 < \Delta t < 1$. At each time step, there are $Q_{ij}\Delta t$ particles per unit time moves from node $i$ to node $j$. Then, the number of particles stored at node $i$ is updated as
\begin{equation}
\label{eqDiscreteParticle}
{\Phi _i}(t + \Delta t) = {\Phi _i}(t) + \sum\limits_{e \in {E_i}} {{Q_e}}
\end{equation}
As for this continuous dynamic system, the process will not stop until the conductivity of each edge between current step and the previous stem is very small. Here, we set $\sum\limits_{i = 1}^n {\sum\limits_{j = 1}^n {\left| {{D_{ij}}(t + \Delta t) - {D_{ij}}(t)} \right|} }  \le {10^{ - 4}}$ is the termination criterion. The implementation detail of the proposed model is given in Algorithm 1.

\begin{algorithm}[H]
\label{algorithm1}
\caption{\textbf{A bio-inspired model for Transshipment Problem}}
\begin{algorithmic}[1]
\STATE // A directed weighted graph $G = (N,E,L)$ has $n$ nodes and $m$ edges. The weight of edge $E_{ij}$ can be denoted as $L_{ij}$.
\STATE \textbf{Initialize:}
    \STATE $S$ is the set of source nodes. $T$ is the set of sink nodes
    \STATE Let ${D_{ij}}: = c~(c > 0)$ and ${Q_{ij}}: = 0$ for each edge;
    \STATE Let the number of particles stored at each node be ${\Phi_i}: = 0$
\WHILE {termination criterion is not met}
    \STATE \textbf{for each} node $i$ in $S$
    \STATE ~~Input $I_{S(i)}$ particles into the source node $S_i$.
    \STATE \textbf{end for}
    \STATE \textbf{for each} node $i$ in $T$
    \STATE ~~\textbf{if} $\Phi_i \ge I_{T(i)}$
    \STATE ~~~~~Output $I_{S(i)}$ particles from the source node $T_i$;
    \STATE~~\textbf{else}
    \STATE~~~~~Output $\Phi_i$ particles from the source node $T_i$
    \STATE~~~\textbf{end if}
    \STATE \textbf{end for}
    \STATE Calculate the flux of each edge $Q_{ij}$ by using Eq.~(\ref{eqofpoiseuille})
    \STATE Update the conductivity of each edge $E_{ij}$ by using Eq.~(\ref{eqDiscreteEvolution})
    \STATE Calculate the number of particles stored at each node by using Eq.~(\ref{eqDiscreteParticle})
    \STATE \textbf{if} $D_{ij} \approx 0$
    \STATE ~~~$D_{ij}: = 0$
    \STATE \textbf{end if}
\ENDWHILE
\STATE Obtain the flux distribution pattern.
\end{algorithmic}
\end{algorithm}

\begin{figure}
\onefigure[scale=0.2]{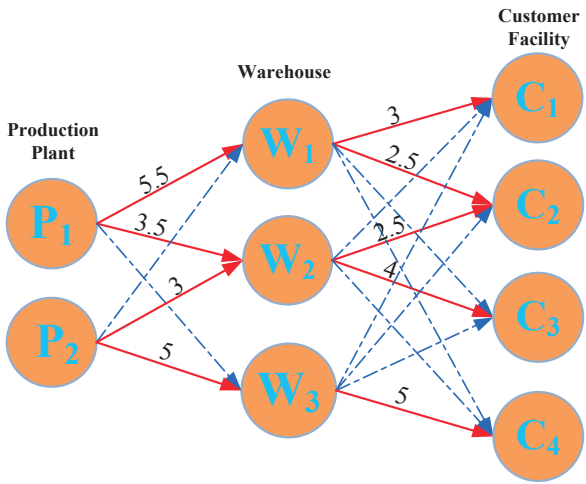}
\caption{A simple transportation network. The direction of the edge limits the movement of products}
\label{fig1}
\end{figure}

\begin{figure}
\centering
  \includegraphics[scale=0.4]{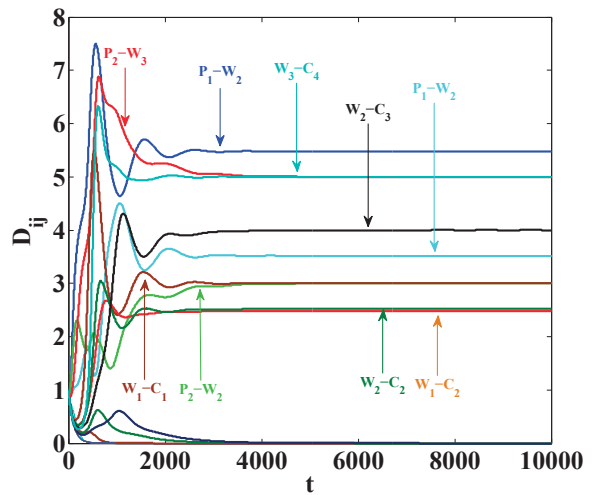}\\
\caption{Superimposed plots of $D_{ij}$'s vs. time where $\Delta t = 0.01$. The initial conductivity of each edge is 1.}
\label{figofConductivity}
\end{figure}
\section{Case study}

\begin{table}[H]
\caption{Cost matrix of a simple example. The cost per unit shipment is given when transferring goods between three nonidentical level.}
\label{table1}
\begin{center}
\begin{tabular}{ccccccc}
\toprule
         & $P_1$ & $P_2$ & $C_1$ & $C_2$ & $C_3$ & $C_4$ \\
\midrule
$W_1$   & 1     & 3     &  5    & 7     & 100  & 100 \\
$W_2$   & 2     & 1     &  9    & 6     & 7    & 100 \\
$W_3$   & 100   & 2     & 100   & 6    & 7     & 4 \\
\bottomrule
\end{tabular}
\end{center}
\end{table}

\begin{table}[H]
\caption{Optimal solution of the given example.}
\label{table2}
\begin{center}
\begin{tabular}{ccccccc}
\toprule
         & $P_1$ & $P_2$ & $C_1$ & $C_2$ & $C_3$ & $C_4$ \\
\midrule
$W_1$   & 5.5     & 0     &  3    & 2.5   & 0  & 0 \\
$W_2$   & 3.5     & 3     &  0    & 2.5   & 4  & 0 \\
$W_3$   & 0       & 5     & 0     & 0     & 0  & 5 \\
\bottomrule
\end{tabular}
\end{center}
\end{table}
As illustrated in Fig.~\ref{fig1}, assume there is company making a certain amount of products at its production plants $P_i~(i=1,2)$. They ship completed the product to warehouse in three different places $W_i~(i=1,2,3)$. Next, they are distributed to customer facilities in four locations $C_i~(i=1,2,3,4)$. The cost matrix is listed in Table~\ref{table1}. The supply of $P_1$ and $P_2$ are 9 and 8, respectively. The demand of $C_i(i=1,2,3,4)$ are 3, 5, 4 and 5, respectively. Note that the total supply is equal to total demand. Our task is to find an optimal shipment plan.

First, we regard the particle flow $Q_{ij}$ as the same amount of goods transported from the sources to the sinks. Consider $P_1$ and $P_2$ as source nodes, $C_i(i=1,2,3,4)$ as sink nodes. According to Algorithm 1, we initialize the conductivity of each edge $D_{ij} = 1$, $\Delta t = 0.01$. As we can see from Fig.~\ref{figofConductivity}, the network is 'empty' without any particle at first iteration. Then, some edges are reinforced along with the pumping particle flow. Finally, the conductivity of each edge converges to a certain number. Some edges among the optimal distribution paths are left with particle flow (draw in distinct colors), while some edges disappear after enough time steps. The flow in each edge is listed in Table~\ref{table2}. Calculate the 'residual' flow along its corresponding edge, we obtain the optimum value $\sum\limits_{e \in {E_r}} {{Q_e}Cos{t_e}}  = 121$, where $E_r$ is the edge set of optimal distribution plan and $Cost_e$ is the cost of edge $e$.
\section{Discussion}
In summary, we present a mathematical model to describe the adaptation process of the given dynamic system. The model can be used to solve transshipment problem, which is regarded as a special multi-source multi-sink minimum cost flow problem. Second, the model has lower computational complexity than \emph{Physarum Solver}. The time complexity of our model is determined by iteration times and the number of edges. In \emph{Physarum Solver}, the time complexity of solving the network Poisson equation is $O({n^3})$ , while our model is $O(m)$, where $m$ is the number of edges in the network. In terms of space complexity, it takes $O(m)$ in our model, generally lower than $O(n^2)$ in \emph{Physarum Solver}. Third, in Ref.~\cite{ito2011convergence}, the authors proposed a non-symmetric update of conductivity modification just like Eq.~(\ref{eqofevolution}). If there are edges in both direction between node $i$ and node $j$ and the flow rate $Q_{ij}$ among $M_{ij}$ is greater than zero, the conductivity $D_{ij}$ will become bigger while $D_{ji}$ is getting smaller. As a consequence, the proposed model can be applied in directed networks as well as undirected ones. Last but not least, the proposed model solves the traditional transshipment problem in a continuous manner. Such a characteristic can be used to deal with complicated problems. For example, due to weather or other unexpected factors, the cost of transferring goods may change dynamically. To the best of our knowledge, the exiting discrete method may recalculate the process back. But in our model, we just keep iterating based the latest exit status (only by using the variable parameter of the conductivity vecter).

\acknowledgments
The work described in this paper is partially supported Chongqing Natural Science Foundation, Grant No. CSCT, 2010BA2003, National Natural Science Foundation of China, Grant Nos. 61174022, 61364030 and 71271061, National High Technology Research and Development Program of China (863 Program) (No.2013AA013801), Science and Technology Planning Project of Guangdong Province, China (2010B010600034, 2012B091100192), Doctor Funding of Southwest University Grant No. SWU110021 and Fundamental Research Funds for the Central Universities No. XDJK2014D008.

\end{document}